\documentstyle[prl,aps,epsf,preprint]{revtex}
\begin{document}
\draft

\title{Supercooled fluid-fluid phase transition in three dimensions
 from a soft-core potential}
\author{
G. Franzese$^1$,
G. Malescio$^2$,
A. Skibinsky$^1$, 
S. V. Buldyrev$^1$, 
and H. E. Stanley$^1$
}

\address{
$^1$Center for Polymer Studies and Department of Physics,
Boston University, Boston, MA 02215, USA\\
$^2$Dipartimento di Fisica, Universit\`a di Messina 
and Istituto Nazionale Fisica della Materia, 98166 Messina, Italy
}
\date{\today}

\maketitle

\begin{abstract}
To study the possibility of a fluid-fluid phase transition, we analyze a
three-dimensional soft-core isotropic potential for a one-component
system.  We utilize two independent numerical approaches, (i) integral
equation in the hypernetted-chain approximation and (ii) molecular
dynamics simulations, and find for both approaches a fluid-fluid phase
transition as well as the conventional gas-liquid critical point.  We
also study the possible existence of a triple point in the supercooled
fluid phase at which three phases---gas, high-density fluid, and
low-density fluid---coexist.  
\end{abstract}



Recent experiments on phosphorus \cite{Katayama} show that a fluid-fluid
phase transition is present at a temperature above the gas-liquid
critical temperature. The transition is between two stable fluids at
different densities. This sort of transition between low-density (LD)
and high-density (HD) fluids has been proposed for many different
materials, including supercooled H$_2$O, Al$_2$O$_3$-Y$_2$O$_3$,
SiO$_2$, GeO$_2$, C, S, Ga, Se, Te, I$_2$, Cs, Bi and Si
\cite{Debenedetti}. Experimental and theoretical studies support this
hypothesis, and some suggest (e.g. for H$_2$O and C) the presence of a
fluid-fluid critical point (CP)
\cite{Katayama,ms,vari,Togaya,Poole,Glosli}.

 Theoretical and numerical 
studies take into account, as accurately as possible, the
anisotropies of the molecular interactions, depending on the relative
molecular orientations \cite{Debenedetti}.  
An alternative approach to the problem consists
of investigating simple isotropic potentials which might be able to give
rise to the LD-HD fluid transition.  This makes it possible to find out
which are the ingredients of the inter-particle interaction related to
LD-HD fluid transition.

Here we study an isotropic potential in three dimensions (3D) using two
methods: (i) integral equations in the hypernetted-chain (HNC)
approximation and (ii) molecular dynamics (MD) simulations.  These
approaches agree qualitatively, showing a phase diagram with a
gas-liquid CP and a LD-HD fluid transition with a second CP.  The
results suggest  the presence of a gas-LD-HD triple point.
The MD simulations reveal
that both transitions are in the supercooled fluid phase, i.e. where the
fluid phase is metastable with respect to the solid.

Stell and Hemmer \cite{hs} proposed the possibility that an isotropic
potential with a region of negative curvature in the repulsive core
(``core-softened potential'') might have two transitions.  Recently it was
shown, using analytic solutions in 1D and MD in 2D, that a Stell-Hemmer
type of interaction gives rise to water-like fluid anomalies that can be
related to the existence of two different local structures in the fluid
phase \cite{ssbs}.  Furthermore, the behavior of spherical particles,
interacting in 3D through an isotropic potential with a hard-core and a
linear repulsive shoulder, has been investigated through Monte Carlo.
\cite{j}.  It was also shown that the addition of an infinite range van
der Waals type attractive term may give rise to a second CP.  The
attraction, however, was not explicitly included in the simulation and
was taken into account in a mean field type scheme \cite{j}.

One feature of LD-HD transition is that there is a competition between
an expanded structure and a compact structure.  In water, for example,
the former is due to hydrogen-bond formation and is preferred at low
pressures and low temperatures, while the latter is favored at high
pressures and high temperatures.  This suggests a potential with two
{\em characteristic radii}
\cite{ms}.  Here we consider a  3D potential (Fig.~\ref{fig1}, inset)
with a hard-core radius $a$ and a soft-core radius $b$.
Depending on the value of pressure and temperature, the effective
hard-core will be $a$ or $b$.  The soft-core is given by a repulsive
square shoulder with height $\lambda \epsilon$.  The attractive part of
the potential is given by a square well, with depth $\epsilon$ and an
interaction cut-off at distance $c$.  Choosing $a$ as length unit and
$\epsilon$ as energy unit, the potential has three free parameters:
$\lambda$ (we set $\epsilon=1$) and ratios $b/a$, $c/a$.  Their possible
combinations are too many to be investigated directly by means of MD.
Therefore, it is advisable to use a faster, way of estimating the phase
diagram.  In this way it is possible to have a first indication of the
phase diagram dependence on the potential's parameters.  This allows one
to perform MD simulation in an {\em systematic} fashion.
 
Integral equations \cite{hm} for the radial distribution function may
provide a convenient tool for estimating the phase stability of the
fluid.  In particular, the HNC equation \cite{hm} has often been used
for relating phase diagrams and potentials \cite{c}.  For this theory
there exists a region in the density-temperature ($\rho$-$T$) plane in
which no solution can be found.  When this region is approached from
high temperatures, there is a strong increase of the isothermal
compressibility $K_T$.  Since there is no genuine divergence in $K_T$, 
this region cannot be identified in a rigorous way with
the spinodal decomposition region of the fluid.  However, for a large variety
of potentials, its shape resembles qualitatively the fluid spinodal
\cite{c}.  This allows us to have a first hint of the topology of the
fluid instability region.

The instability line, below which no solution of the HNC can be found,
is shown in Figs.~\ref{fig1}, \ref{fig2}.  For $b/a=1.4$, $c/a=1.7$ and
$\lambda=-0.5$, the potential considered in Ref.~\cite{ssbs}, the shape
of the instability line is similar to that of a standard gas-liquid
spinodals and does not change significantly by varying the value of
$\lambda$.  However, for other combinations of the parameters, the shape
can change dramatically (Fig.~\ref{fig2}).  For example, for $b/a=2$ and
$c/a=2.2$, we have a bifurcated instability line at intermediate values
of $\lambda$.  The HNC instability line is not a true spinodal, but its
bifurcation, with two distinct maxima, supports the existence of two CPs
\cite{note2}.  Note that the bifurcation in the instability line
disappears by increasing $\lambda$ (as shown in Fig.~\ref{fig2}.A) or by
changing only one of the other parameters.  Therefore, the phase diagram
depends drastically on the delicate interplay between the parameters.

Based on the HNC results, we performed extensive MD simulations for a
system of particles interacting through the potential with $b/a=2.0$,
$c/a=2.2$ and $\lambda=0.5$ \cite{note3}.  At high density the stable
phase is solid, but the MD allows us to study the supercooled fluid
phase before the crystallization occurs.  The life-time of this
metastable phase depends on ($\rho$-$T$).  To be sure that our estimates
are done in the fluid phase, we study the structure factor
$S(\vec{Q},t)\equiv [\rho_{\vec{Q}}^*(t)\rho_{\vec{Q}}(t)]/N$ where
$\rho_{\vec{Q}}\equiv \sum_j \exp(i\vec{Q}\cdot\vec{r}_j(t))$, $N$ is
the number of particles, $\vec{Q}$ is the wave vector
\cite{notek} and
$\vec{r}_j(t)$ is the position of particle $j$ at time $t$.  The growth
of $S(\vec{Q},t)$ for wave vectors with $Q\rightarrow 0$ indicates a
phase separation.  Furthermore, for finite $Q$, it marks the onset of
the crystal phase.  Our simulations show the occurrence of a phase
separation without a crystal formation (time interval labeled with B in
Fig.~\ref{fluct}), followed by the onset of crystal phase (C in
Fig.~\ref{fluct}) marked by a dramatic increase of $S(\vec{Q},t)$ for
$Q\simeq 6$ and 12 (in $a^{-1}$ units) \cite{noteEP}.  Our calculations are
done in the (metastable) fluid phase before the crystallization.
 
The study of the static structure factor $S(Q)\equiv [\langle S(Q,t)
\rangle]$, where $\langle \cdot \rangle$ is the average over $\vec{Q}$
vectors, and $[\cdot]$ is the average over a time interval, done on
different time intervals (A, B and C in Fig.~\ref{fluct}), confirms the
previous analysis. To emphasize the phase separation occurring in the
fluid phase, we show in Fig.~\ref{sep} the snapshot corresponding to
the larger peak in time-interval B in Fig.~\ref{fluct}.a. The histograms
of the number of particles in the system for the three planar
projections of the snapshot show a clear separation in density.
Therefore, the phase separation marked by the increasing of $S(Q)$ at
low values of $Q$ is associated with the coexistence of LD and HD
fluids.

The MD phase diagram is shown in Fig.~\ref{fig3}.  At low densities and
high temperature, the system is in the gas phase. Increasing the density
or decreasing the temperature, the system undergoes a first order
gas-solid transition.  In the supercooled fluid phase, we observe two
regions with negatively-sloped isotherms.  The first region is for
temperatures $k_BT/\epsilon <0.61 $ in the low density regime.  The
second region is for temperatures $k_BT/\epsilon <0.645 $ in the high
density regime.  These regions correspond to thermodynamically unstable
states and their borders are the spinodal lines.
In the region of low density the coexisting fluids are the gas and the
liquid.  In the region of high density the analysis presented above
allows one to conclude that the coexisting phases are a LD and a HD
fluid.  The points at highest pressure on the spinodal lines correspond
to the CPs.
In both HNC and MD approaches the merging of instability regions below
the critical temperatures suggests the merging of coexisting lines, i.e.
the existence of a gas-liquid-liquid triple point.

Within our precision we find that the gas-liquid critical point is at
$k_BT_{c1}/\epsilon =0.605 \pm 0.005 $, $a^3 P_{c1}/\epsilon =0.018 \pm
0.002 $ and $a^3 \rho_{c1}/m= 0.115 \pm 0.015$ corresponding to a
packing fraction $\eta_{c1}=0.06\pm0.01$ ($\eta=\pi a^3\rho/6$).  Note
that $T_{c1}$ and $\rho_{c1}$ are in good agreement with the HNC
results.  The second instability region occurs at packing fraction $
\eta > 0.08$, deep in the solid phase, where the metastable fluids
phases have short lifetimes.
Within our precision the critical temperature is $k_BT_{c2}/\epsilon =
0.650\pm0.005 $, the critical pressure is $a^3 P_{c2}/\epsilon=0.08 \pm
0.01$ and the critical packing fraction is $\eta_{c2}=0.16\pm 0.01$.
Note that $T_{c2}$ is consistent with the HNC result, while $\rho_{c2}$
is smaller.

Since the MD calculations show that $T_{c2}>T_{c1}$, the above set of
parameters is not able to reproduce a fluid-fluid critical point.
Nevertheless the HNC analysis suggests that other sets of parameters can
lead to $T_{c2}<T_{c1}$ as shown in Fig.~\ref{fig2}.b.
Preliminary calculation on a second set of parameters with $b/a=2.2$,
$c/a=2.4$ and $\lambda=0.5$ show that both CPs are lowered in
temperature and pressure.

In conclusion, we studied an isotropic soft-core potential with a
repulsive shoulder and an attractive well by means of integral equations
in the hypernetted-chain approximation (HNC) and molecular dynamics (MD)
simulations. 
Both approaches 
show, in the supercooled fluid phase, a gas-liquid
critical point (at low density, low pressure, low temperature) and a 
transition (at high density, high pressure, high temperature) 
between a low-density (LD) and a high-density (HD) fluid with a second
critical point. 
This second transition resembles the fluid-fluid transition discovered
recently in experiments with phosphorus \cite{Katayama}.
However, in phosphorus the fluid-fluid transition occurs
between stable phases, while in our case it is in the supercooled phase
as in H$_2$O. 
Furthermore, both numerical approaches used here 
suggest the existence of a
gas-liquid-liquid triple point. 
Therefore, the potential proposed here, despite its
simplicity, is able to reproduce many intriguing features of materials
with a second fluid-fluid transition, leading to a new way to explore
the existence of a second critical point.

We wish to thank 
P.V. Giaquinta, 
G. Pellicane, 
A. Scala,
F. Sciortino,
F. W. Starr, 
for helpful suggestions and for interesting and stimulating discussions. 
We thank NSF for support and CNR grant N.203.15.8/204.4607
(Italy) for partial support.



\begin{figure}
\caption{ Instability line in ($\rho$-$T$) plane of the HNC equation for
$b/a=1.4$, $c/a=1.7$ $\epsilon=1$ and $\lambda$ given by the label of
the curves. Inset: general form of the soft-core potential.
} 
\label{fig1}
\end{figure}

\begin{figure}
\caption{As in Fig.1 with 
(A) $b/a=2.0$, $c/a=2.2$ and 
(B) $b/a=2.5$, $c/a=3.0$ and 
$\lambda$ given by the label of the curves.
}
\label{fig2}
\end{figure}

\begin{figure}
\caption{
Structure factor versus time for (a) wave vectors with $Q\simeq 1$ and
(b) $Q\simeq 12$ (in $a^{-1}$ units) 
at density $a^3\rho/m=0.27$
and $k_BT/\epsilon=0.62$ up to $3\times 10^4$ UPP; 
(c) The static structure factor $S(Q)$ averaged over
different time intervals: labels (A,B,C) stands for the time intervals
shown in panels (a,b). For A (B) we averaged over a low (high) density
fluctuation time interval; for C over an interval after the crystal seed
formation. The curves for 
$S(Q)$ are shifted for the sake of comparison. 
}
\label{fluct}
\end{figure}

\begin{figure}
\caption{
Histograms of number of particles as function of one coordinate for the
snapshot corresponding to the large value of $S(Q=1,t)$ in time interval
B of previous figure. The corresponding projections of the snapshot are
also shown. Histogram and snapshots are shifted for the sake of clarity. 
The horizontal line shows the average density.
}
\label{sep}
\end{figure}

\begin{figure}
\caption{Pressure-density isotherms from the MD simulation for the
potential with $b/a=2.0$, $c/a=2.2$ and $\lambda=0.5$. Full points
represent stable fluid. Open points represent supercooled
fluid \protect\cite{note3}. The inset emphasizes the instability region
with negatively-sloped isotherms (and a van der Waals loops at
$k_BT/\epsilon=0.6$) associated with the gas-liquid CP. It is possible to
see a high density region with the van der Waals loops associated with
the second CP. 
The lines connecting the points are just 
guides for the eyes.
Temperature is measured in $\epsilon/k_B$.
}
\label{fig3}
\end{figure}

\begin{figure}
\mbox{ \epsfxsize=18cm \epsffile{ 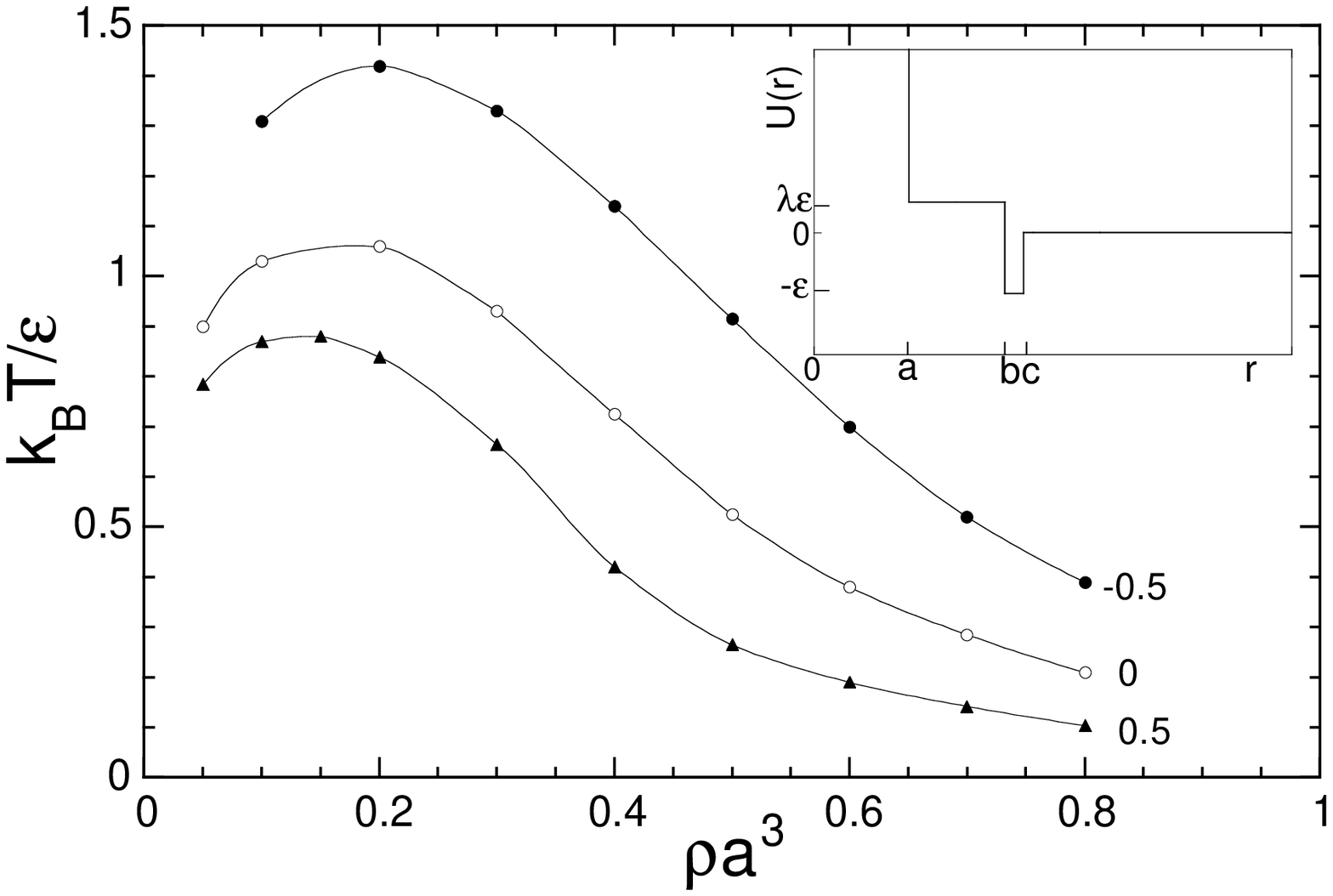 } }
\end{figure}

\begin{figure}
\mbox{ \epsfxsize=18cm \epsffile{ 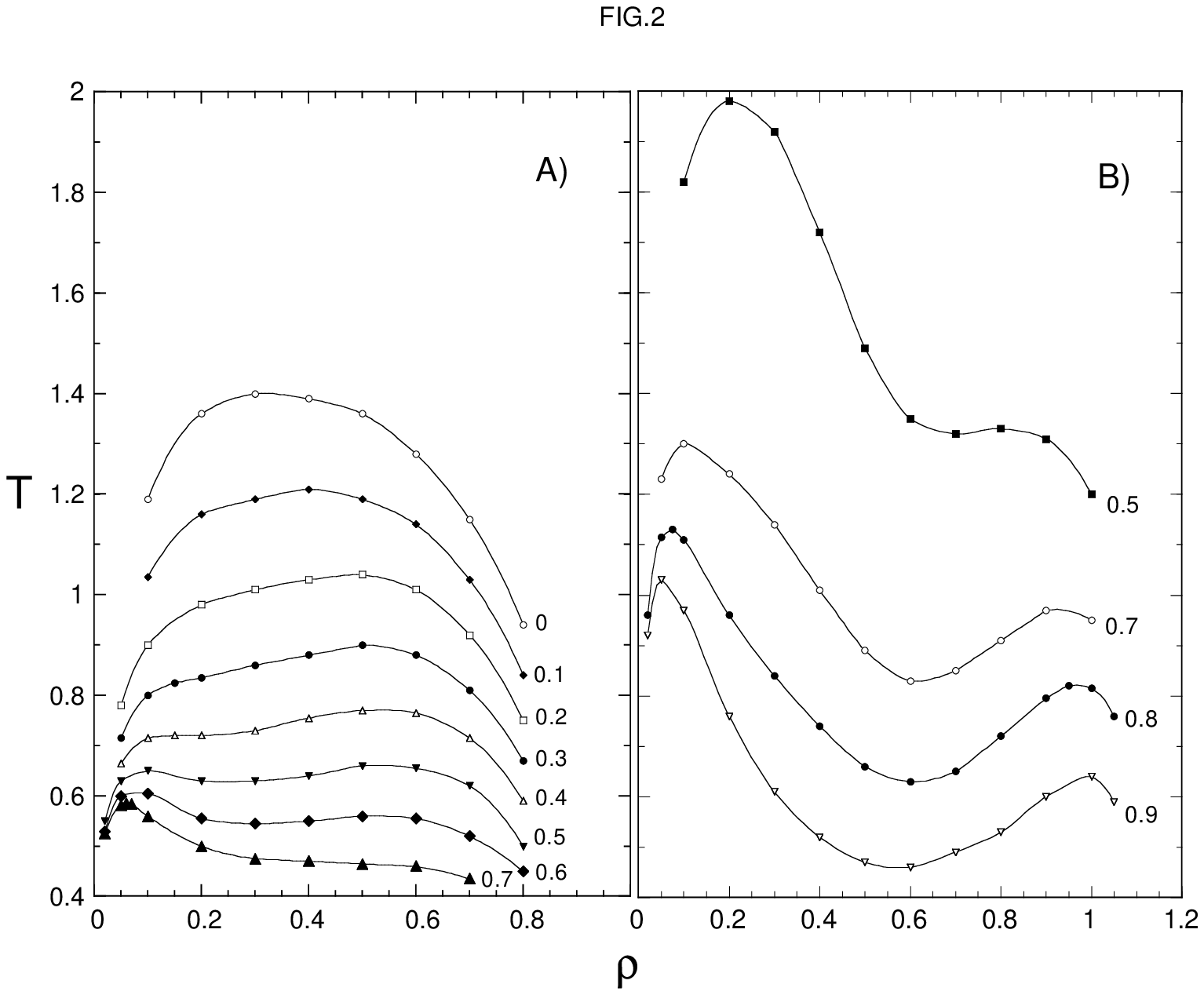 } }
\end{figure}

\begin{figure}
\mbox{ \epsfxsize=14cm \epsffile{ 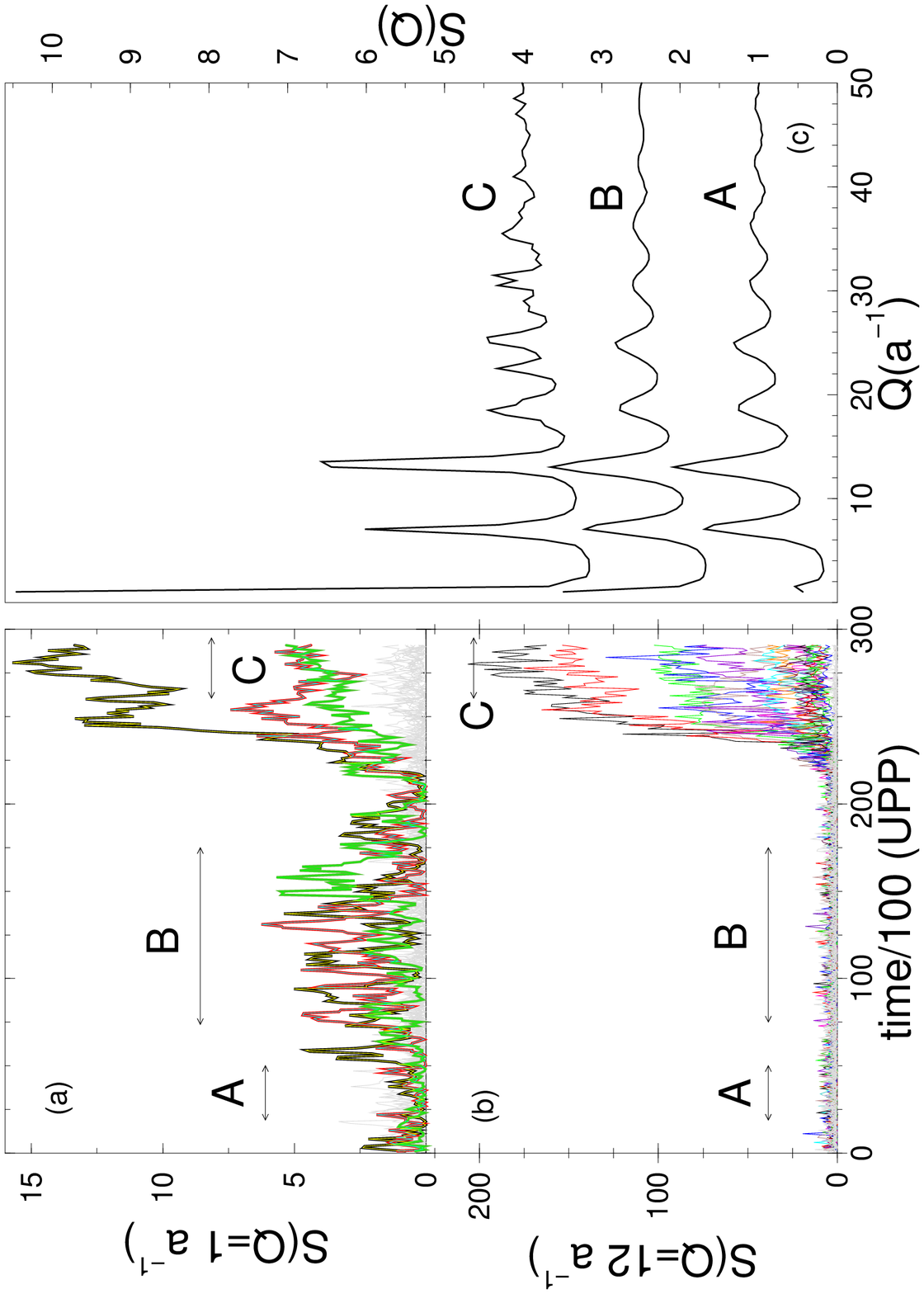 } }
\end{figure}

\begin{figure}
\mbox{ \epsfxsize=14cm \epsffile{ 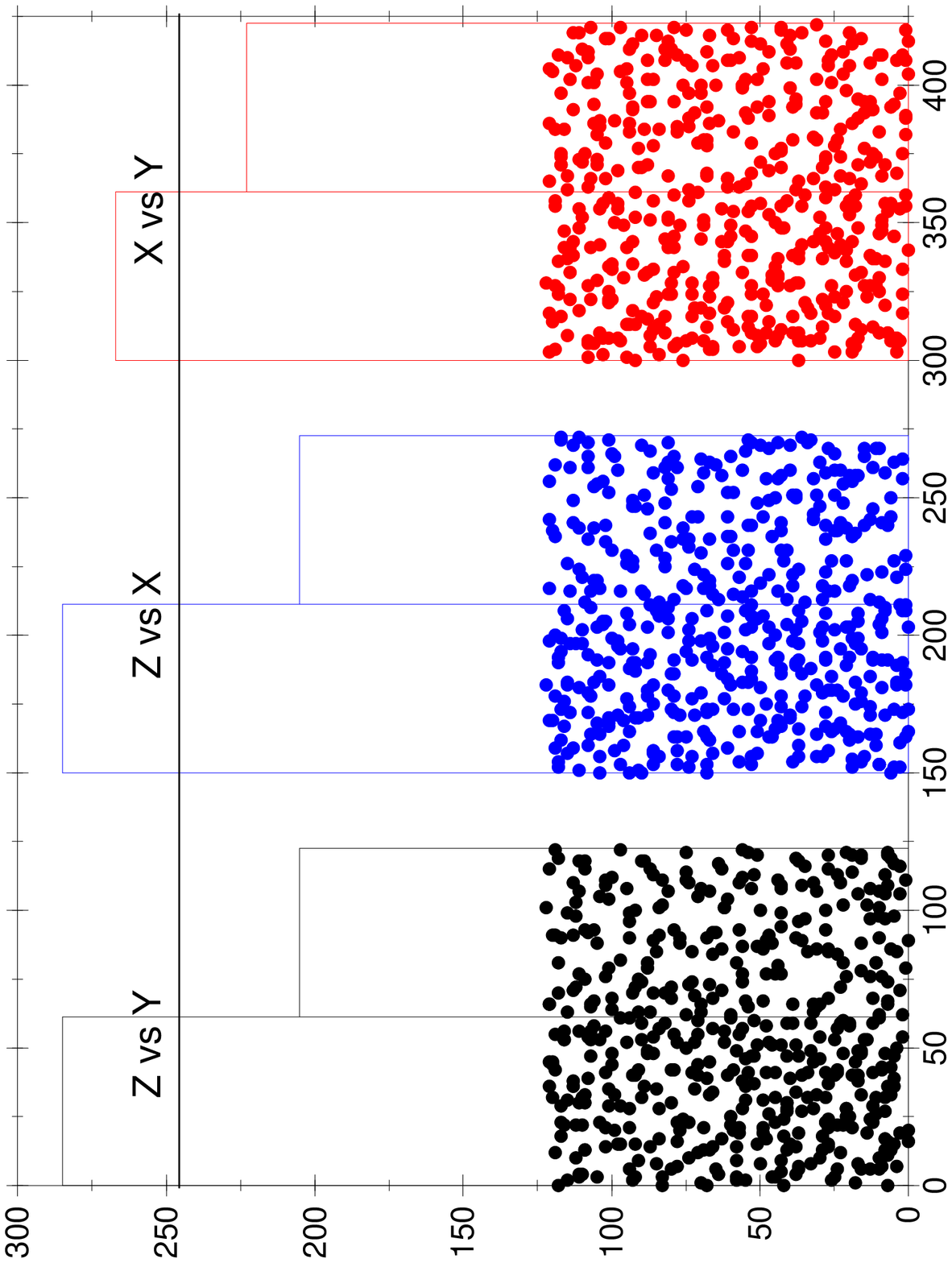 } }
\end{figure}

\begin{figure}
\mbox{ \epsfxsize=14cm  \epsffile{ 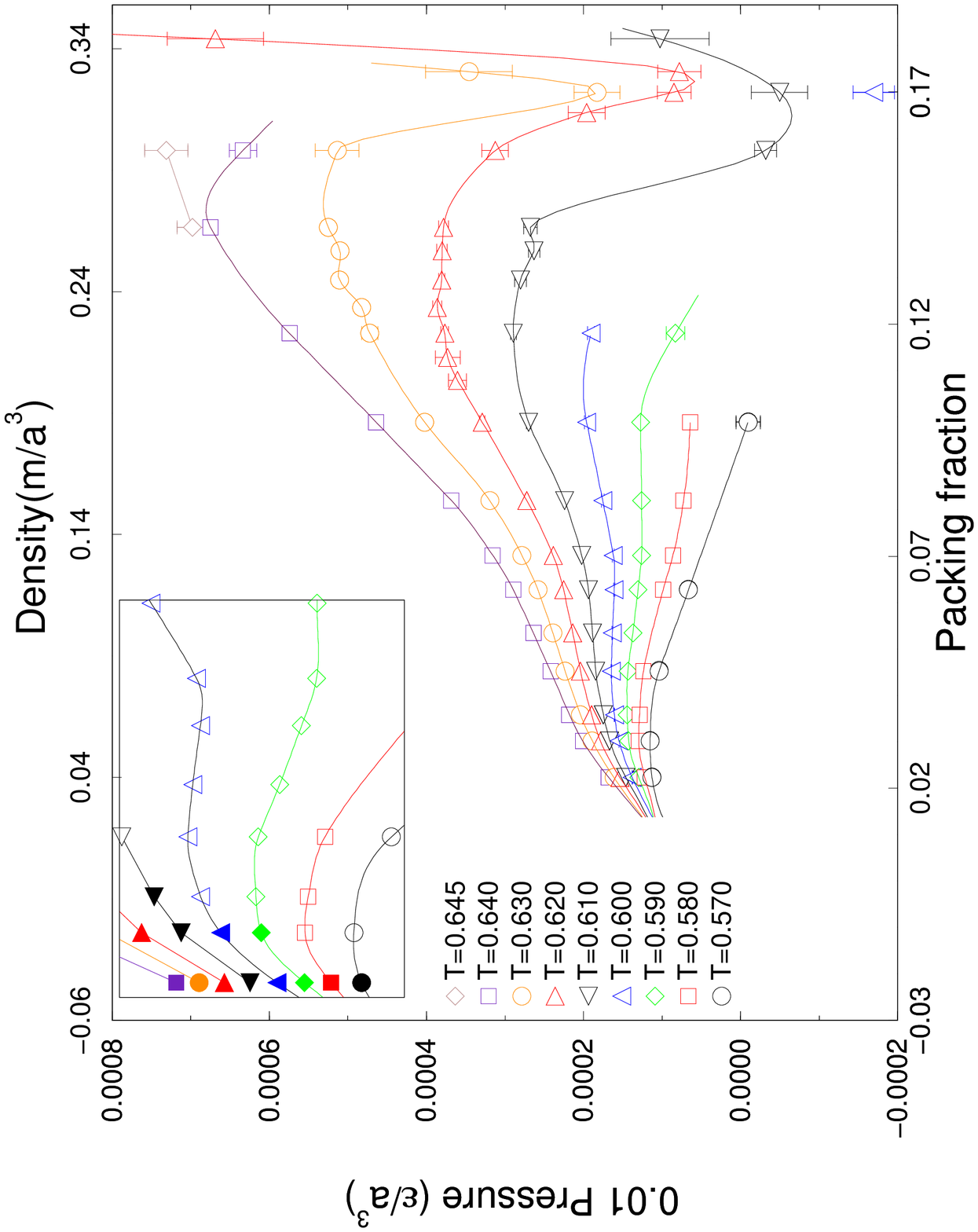 } }
\end{figure}


\begin{thebibliography}{99}
\bibitem{note2}
We observe that the overall shape of the bifurcated instability line is consistent 
with the phase diagram derived by Tejero and Baus \cite{tb} within a van der Waals 
approach for an effective density-dependent potential.
\bibitem{note3}
The MD simulations are performed fixing the number of particles $N=490$,
the volume $V$ and the average temperature $k_BT/\epsilon$ ($k_B$ is the
Boltzmann constant).
We use a standard collision event list algorithm \cite{r} to evolve the
system and a modified Berendsen method to achieve the desired
temperature \cite{b}. 
To study the equilibrium phase diagram for each $V$ we equilibrate the
system at low enough temperature (e.g. $k_BT/\epsilon=5$ for
$a^3\rho/m=0.04$ with $m$ unitary mass) 
to form a crystal seed in the liquid phase. 
Starting from this configuration, we perform simulations at different
$T$ to see which phase prevails and to find the phase separation line. 
To study the (metastable) fluid phase at each $V$ 
we quench the system  starting from a high
$T$ configuration. 
We observe that, in the ($\rho$-$T$) region considered here, 
$\Delta t= 2\times 10^3$ updating per particle (UPP) are enough on
average to equilibrate the system toward a stable (metastable) 
state in the (supercooled) fluid phase. 
Therefore we discarded the first $\Delta t$ UPP of any simulation and
averaged each data point over at least $10^5$ - $10^6$ UPP.
\bibitem{notek}
We consider up to $9\times 10^4$ wave vectors $\vec{Q}$ with
$|\vec{Q}|\le 100~a^{-1}$.
\bibitem{noteEP}
We tested that  corresponding to the large increase in $S(\vec{Q},t)$,
there is also a dramatic
step-like decrease (increase) of energy (pressure).
\end{thebibliography}
\end{document}